\documentclass[lettersize,journal]{IEEEtran}
\usepackage{amsmath,amsfonts}
\usepackage{algorithmic}
\usepackage{algorithm}
\usepackage{array}
\usepackage[caption=false,font=normalsize,labelfont=sf,textfont=sf]{subfig}
\usepackage{textcomp}
\usepackage{stfloats}
\usepackage{url}
\usepackage{verbatim}
\usepackage{graphicx}
\usepackage{cite}
\usepackage{authblk}
\usepackage{adjustbox}
\usepackage{multirow}
\usepackage{makecell}
\usepackage{tabularx}
\usepackage{booktabs}
\usepackage{tikz}
\usepackage{mathcomSTEv4}
\usepackage{enumitem}
\usepackage{subfig} 
\usetikzlibrary{shapes, arrows.meta, positioning, external, calc, shapes.geometric}
\usepackage{pgfplots}
\pgfplotsset{compat=1.17}
\begin{document}

\title{HAAQI-Net: A Non-intrusive Neural Music Audio Quality Assessment Model for Hearing Aids}

\author{Dyah A. M. G. Wisnu,~\IEEEmembership{Student Member,~IEEE}, Stefano Rini,~\IEEEmembership{Member,~IEEE}, Ryandhimas E. Zezario,~\IEEEmembership{Member,~IEEE}, Hsin-Min Wang,~\IEEEmembership{Senior Member,~IEEE}, and Yu Tsao,~\IEEEmembership{Senior Member,~IEEE}
\thanks{Dyah A. M. G. Wisnu is with the Taiwan International Graduate Program—Social Network and Human Centered Computing, Institute of Information Science, Academia Sinica, Taipei, Taiwan, and also with the College of Informatics, National Chengchi University, Taipei, Taiwan (e-mail: dyahayumgw@iis.sinica.edu.tw)}
\thanks{Stefano Rini is with Institute of Communications Engineering, National Yang Ming Chiao Tung University, Hsinchu, Taiwan (email: stefano@nctu.edu.tw)}
\thanks{Ryandhimas E. Zezario is with the Research Center for Information Technology Innovation, Academia Sinica, Taipei, Taiwan (email: ryandhimas@citi.sinica.edu.tw)}
\thanks{Hsin-Min Wang is with the Institute of Information Science, Academia Sinica, Taipei, Taiwan (email: whm@iis.sinica.edu.tw)}
\thanks{Yu Tsao is with the Research Center for Information Technology Innovation, Academia Sinica, Taipei, Taiwan (e-mail:
yu.tsao@citi.sinica.edu.tw)}}

\maketitle

\begin{abstract}

This paper introduces HAAQI-Net, a non-intrusive deep learning-based music audio quality assessment model for hearing aid users. Unlike traditional methods like the Hearing Aid Audio Quality Index (HAAQI) that require intrusive reference signal comparisons, HAAQI-Net offers a more accessible and computationally efficient alternative. By utilizing a Bidirectional Long Short-Term Memory (BLSTM) architecture with attention mechanisms and features extracted from the pre-trained BEATs model, it can predict HAAQI scores directly from music audio clips and hearing loss patterns.
Experimental results demonstrate HAAQI-Net's effectiveness, achieving a Linear Correlation Coefficient (LCC) of $0.9368$, a Spearman's Rank Correlation Coefficient (SRCC) of $0.9486$, and a Mean Squared Error (MSE) of $0.0064$ and inference time significantly reduces from $62.52$ to $2.54$ seconds. To address computational overhead, a knowledge distillation strategy was applied, reducing parameters by $75.85$\% and inference time by $96.46$\%, while maintaining strong performance (LCC: $0.9071$, SRCC: $0.9307$, MSE: $0.0091$).
To expand its capabilities, HAAQI-Net was adapted to predict subjective human scores like the Mean Opinion Score (MOS) through fine-tuning. This adaptation significantly improved prediction accuracy, validated through statistical analysis. Furthermore, the robustness of HAAQI-Net was evaluated under varying Sound Pressure Level (SPL) conditions, revealing optimal performance at a reference SPL of $65$ dB, with accuracy gradually decreasing as SPL deviated from this point.
The advancements in subjective score prediction, SPL robustness, and computational efficiency position HAAQI-Net as a scalable solution for music audio quality assessment in hearing aid applications, contributing to efficient and accurate models in audio signal processing and hearing aid technology.
\end{abstract}


\section{Introduction}
\IEEEPARstart{D}{espite} advances in audio processing technology, music perception remains a complex challenge for individuals using hearing aids \cite{leal2003music} as hearing aids are tailored to the specific hearing loss characteristics of each user \cite{edwards2007future}. 
This challenge highlights the need for effective assessment methods, which can be broadly categorized into subjective and objective approaches. While subjective research methods, such as listening tests, provide valuable insights into how individuals perceive music and sound quality, they are resource-intensive, often requiring large sample sizes and controlled environments. Furthermore, subjective results can be influenced by individual biases, variability in listeners' auditory perception, and factors such as emotional state or cognitive load, which can complicate the generalization of findings. In contrast, objective metrics, while offering consistency and scalability, may not fully capture the subtleties of human perception and can sometimes diverge from the subjective experience of sound quality.

Numerical methods such as Mean Squared Error (MSE), Signal-to-Noise Ratio (SNR), and Signal-to-Distortion Ratio (SDR), while providing quantitative metrics, often fail to align closely with human auditory perception \cite{vincent2006performance}. The Hearing Aid Audio Quality Index (HAAQI) \cite{kates2015hearing} provides a music quality score that is more congruent with human perception. However, HAAQI has several limitations: (i) it requires the availability of ground truth signals, which makes it unusable in situations where such signals are unavailable; (ii) its computational complexity restricts its applicability, especially in real-time processing scenarios or low-resource environments; (iii) since it is non-differentiable, it cannot be directly incorporated into deep learning models for downstream applications as a loss function. These properties limit its applicability in practical scenarios, such as when an objective assessment must be made on the fly, or when it is used as a loss function in training deep learning models for music audio enhancement.

In this paper, we propose HAAQI-Net, a deep learning-based neural counterpart of HAAQI that overcomes the above limitations of HAAQI. HAAQI-Net is designed to be (i) non-intrusive, enabling assessment without the need for ground truth signals; (ii) efficient, allowing real-time processing and reducing computational overhead; and (iii) differentiable, facilitating its integration into deep learning frameworks for training and optimization. By addressing these challenges, HAAQI-Net provides a promising solution for objective music audio quality assessment for hearing aid users, offering improved accessibility, efficiency, and compatibility with modern computational frameworks \cite{kates2015hearing}. This innovation has huge potential to enhance the overall listening experience for people with hearing loss, helping to improve quality of life and participation in social activities \cite{edwards2007future}.

The proposed HAAQI-Net model involves a Bidirectional Long Short-Term Memory (BLSTM) model followed by an attention mechanism. We use the pre-trained BEATs model \cite{BEATs} as the feature extractor. Although HAAQI-Net has high performance, feature extraction through the large BEATs model results in a certain computational overhead.  
Therefore, we reduce the number of parameters in the HAAQI-Net architecture by incorporating knowledge distillation to transfer expertise from a large teacher model (HAAQI-Net with BEATs) to a compact student model (HAAQI-Net with distillBEATs).
Furthermore, we also consider an adaptive distillation strategy to dynamically adjust the loss weight of each training sample based on its difficulty, enhancing the student model's learning process. 
These extensions improve the efficiency and effectiveness of HAAQI-Net in real-world applications, particularly in scenarios requiring real-time processing or low-latency inference.
The effectiveness of HAAQI-Net  is investigated through extensive numerical experimentation.
When compared with the true HAAQI scores, the predicted scores have a Linear Correlation Coefficient (LCC) of $0.9368$, a Spearman's Rank Correlation Coefficient (SRCC) of $0.9486$, and a Mean Squared Error (MSE) of $0.0064$, and the inference time is significantly reduced from $62.52$ seconds to $2.54$ seconds.
Furthermore, the distilled HAAQI-Net not only maintains high-quality predictions but also significantly reduces the runtime under different settings. 
This improvement makes HAAQI-Net more useful in practical applications of music audio quality assessment for hearing aid users, where efficient processing is crucial for providing timely and accurate feedback to users.

While HAAQI-Net represents a meaningful contribution, it is essential to consider this work within the broader context of audio quality assessment methods. In this context, although there are many speech assessment metrics \cite{cooper2024review}, such as Mean Opinion Score (MOS) \cite{itu1996800}, Perceptual Evaluation of Speech Quality (PESQ) \cite{recommendation2001perceptual}, Perceptual Objective Listening Quality Analysis (POLQA) \cite{polqa_2013, brachmanski2022subjective}, Speech Transmission Index (STI) \cite{ref_38}, Normalized-Covariance Measure (NCM) \cite{ncm}, Short-Time Objective Intelligibility (STOI) \cite{taal2011algorithm}, extended STOI (eSTOI) \cite{estoi}, Spectrogram Orthogonal Polynomial Measure (SOPM) \cite{somr}, Neurogram Orthogonal Polynomial Measure (NOPM) \cite{nopm}, and Neurogram Similarity Index Measure (NSIM) \cite{nsim}, there are relatively few dedicated measures for music audio quality assessment, especially for hearing aids. 
Music (and various other audio) quality assessment is broadly divided into intrusive and non-intrusive methods. Intrusive methods involve comparing a corrupted or processed signal to be evaluated with the original signal. Common intrusive methods include Perceptual Evaluation of Audio Quality (PEAQ) \cite{thiede2000peaq}, PEMO-Q \cite{huber2006pemo}, and HAAQI \cite{kates2015hearing}. PEAQ and PEMO-Q do not take hearing loss into account, whereas HAAQI is designed to predict music quality for hearing aid users. While compensating for hearing loss, hearing aids pose distinct challenges, such as degradation of sound quality due to factors such as nonlinear processing and amplification. HAAQI utilizes an auditory model attuned to impaired hearing. It then assesses the quality by comparing the outputs of this auditory model for both the degraded signal and the reference signal. By evaluating differences in signal characteristics, such as envelope modulation and temporal fine structure, HAAQI addresses challenges associated with background noise, nonlinear processing, and varied listening environments of hearing aid users. Traditional non-intrusive assessment methods of audio quality include 3SQM and ITU-T Recommendation P.563 \cite{8937202}. A Learning-to-Rank (LTR) method for music audio quality assessment is proposed in \cite{li2013non}. Recently, non-intrusive neural models for speech assessment have been proposed based on deep learning architectures, such as BLSTM \cite{fu2018quality}, Convolutional Neural Network (CNN) \cite{mittag2021nisqa}, CNN-BLSTM \cite{zezario2022deep}, and Transformer \cite{vaswani2017attention}. Meanwhile, many studies have also focused on developing speech assessment models for hearing aid users \cite{chiang2021hasa,tu22_interspeech, edozezario22_interspeech, MAWALIM2023109663, cuervo2024speech, mogridge2024nonintrusive}. However, to the best of our knowledge, there are no neural models specifically designed for non-intrusive music audio quality assessment for hearing aid users. This gap in the literature motivates the development of HAAQI-Net, a deep learning-based neural counterpart of HAAQI, which aims to address the need for efficient and accurate non-intrusive music audio quality assessment methods tailored for hearing aid users. 

The remainder of this paper is organized as follows. Section II illustrates the proposed methodology. Section III outlines the experimental setup and reports the results. Finally, Section IV provides conclusions and discusses future work. 

\section{HAAQI-Net}
\label{sec:haaqi_net}
This section first introduces the BEATs model for feature extraction, then the model architecture and training objective of the proposed HAAQI-Net model, and finally the knowledge distillation strategy to improve the efficiency of HAAQI-Net.

\subsection{BEATs}
\label{sec:BEATS}
Bidirectional Encoder representation from Audio Transformers (BEATs) \cite{BEATs} is a pre-trained framework that addresses key challenges in audio processing with its innovative acoustic tokenizer and audio Self-Supervised Learning (SSL) model.
This framework iteratively enhances performance by using an acoustic tokenizer for generating discrete labels from unlabeled audio and optimizing the audio SSL model with masking and a discrete label prediction loss. Moreover, this audio SSL model achieves state-of-the-art performance on a variety of audio classification benchmarks, surpassing predecessors that utilize broader training data and a larger number of model parameters \cite{BEATs}.

In this work, we utilize BEATs as the feature extractor. The process starts with pre-processing to extract Mel-Frequency Cepstral Coefficients (MFCCs). Then, it employs patch-based embedding through a convolutional layer, followed by normalization. 
Processed through a Transformer encoder, these embedded features are adept at capturing both local and global dependencies. The whole process is expressed as:
\begin{align}
\text{\textbf{X}}_{\text{BEATs}}^i = \text{\textit{TE}}\left(\text{\textit{LN}}\left(\text{\textit{PE}}\left(\text{\textit{Prep}}(\text{\textbf{X}})\right)\right)\right),
\end{align}
where \textit{TE} denotes the Transformer encoder, \textit{LN} is layer normalization; \textit{PE} stands for patch embedding; \textit{Prep} represents the pre-processing operation; \text{\textbf{X}} is the input waveform; and $\text{\textbf{X}}_{\text{BEATs}}^i$ is the output features from the \textit{i}-th layer of BEATs' Transformer encoder.

To ensure we capture detailed information from all layers of BEATs' Transformer encoder, we utilize the outputs of all layers instead of just the last one. This approach prevents the loss of nuanced details present in the earlier layers. By extracting and incorporating the outputs of all layers, we can comprehensively grasp meaningful information across the entire sequence. We apply a weighted sum operation to these outputs, allowing us to blend them effectively and ensure that no valuable information is overlooked, which is expressed as:
\begin{align}
\textbf{X}_\text{w\_sum} &= \sum_{i=1}^{L} \left( \text{\textit{LN}}(\text{\textbf{X}}_{\text{BEATs}}^i) \times \sigma(w_i) \right),
\end{align}
where $L$ is the number of layers in the Transformer encoder; $\sigma$ denotes the softmax activation function; and \( w_i \) is the learnable weight associated with $\text{\textbf{X}}_{\text{BEATs}}^i$. The element-wise multiplication of the normalized BEATs features and their respective weights ensures that more attention is given to certain parts of the input features during processing. Finally, the weighted features are summed together to obtain the final input feature vector, \( \textbf{X}_\text{w\_sum} \).

\subsection{Network Architecture}
\label{subsec:network_architecture}

\begin{figure}[t]
    \centering
    \includegraphics[width=1.0\linewidth]{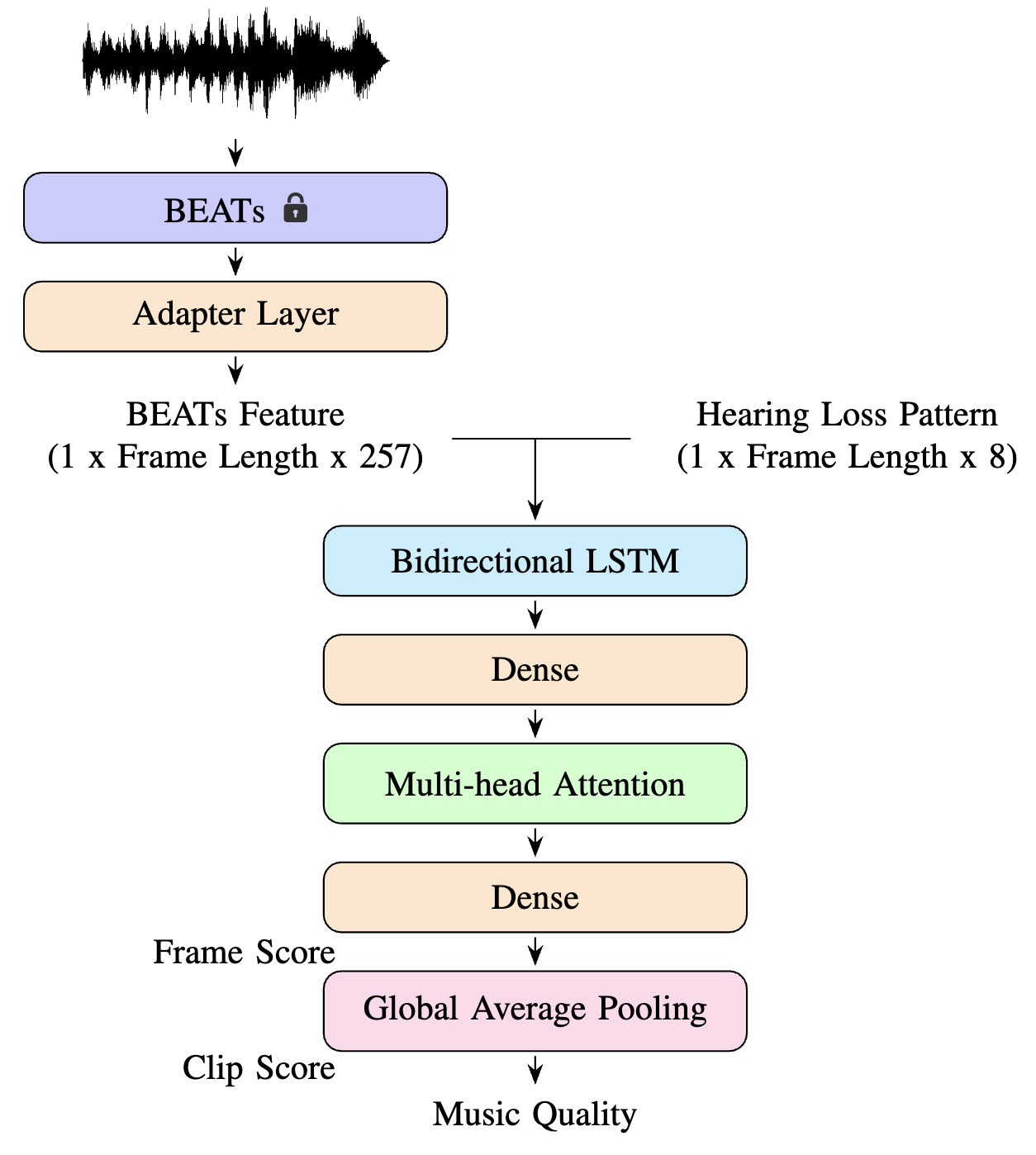}
    \caption{The architecture of HAAQI-Net.}
    \label{fig:system_architecture}
\end{figure}

The overall architecture of HAAQI-Net is illustrated in Fig. \ref{fig:system_architecture}. The input waveform is passed through the pre-trained BEATs model to obtain audio features. 
An adapter layer implemented as a dense layer is used to adapt these features specifically for music audio quality assessment. This adaptation enhances the model's ability to learn more compact and salient feature representations and capture intricate relationships in the data.
The input to HAAQI-Net includes the adapted BEATs features and the hearing loss pattern. 
The core of HAAQI-Net consists of a BLSTM layer that captures the unique time-varying characteristics of music signals. 
The BLSTM layer is followed by a fully connected layer with $256$ nodes activated by the Rectified Linear Unit (ReLU). 
A multi-head attention mechanism with $16$ heads captures temporal dependencies in the data.
A linear layer followed by a sigmoid activation produces frame-level scores. 
The frame-level scores are then aggregated by a global average pooling layer to form an overall clip quality assessment.

In the training phase, the model processes input tensors with dimensions $[B, T, F]$, where $B$ denotes the batch size, $T$ is the frame number per training music clip, and $F$ is the dimension of the feature vector. 
The objective function for training HAAQI-Net is the sum of the clip-level loss and the averaged frame-level loss:
\begin{align}
L_{Qual} = \frac{1}{B}\sum_{n=1}^{B} \left[ (\hat{Q}_n - Q_n)^2 + \frac{1}{T_n}\sum_{t=1}^{T_n}(\hat{Q}_n - q_{n,t})^2 \right],
\end{align}
where $\hat{Q}_n$ and $Q_n$ represent the true and estimated clip-level quality scores for the $n$-th training music clip, respectively; $T_n$ is the number of frames in the $n$-th training music clip, and $q_{n,t}$ denotes the estimated frame-level quality score of the $t$-th frame of the $n$-th training music clip. During training, the pre-trained BEATs model is fixed.

\begin{figure}[t]
    \centering
    \includegraphics[width=\linewidth]{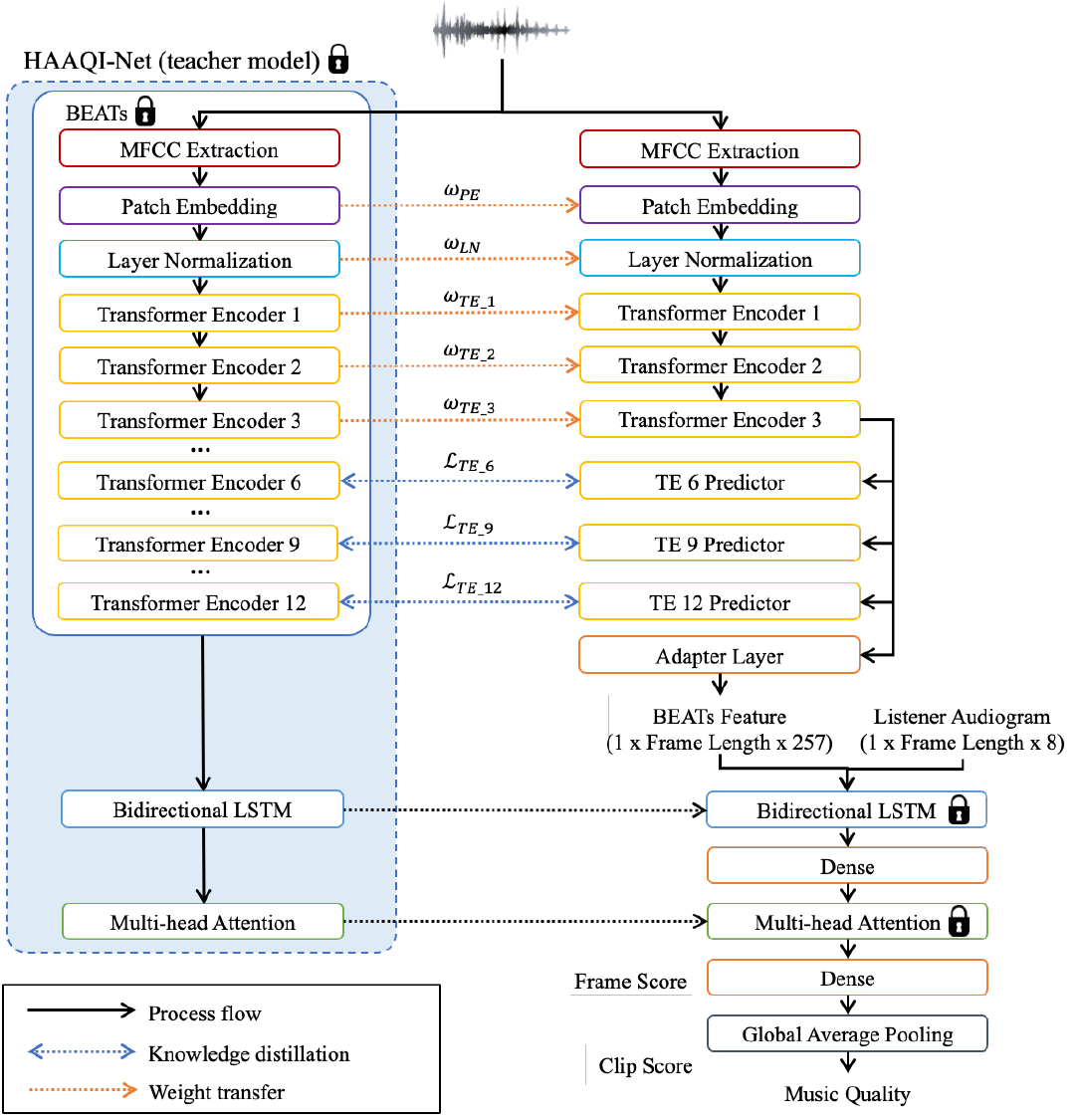}
\caption{The architecture of HAAQI-Net with knowledge distillation.}
  \label{fig:distillHAAQI}
\end{figure}

\subsection{Knowledge Distillation}

In our HAAQI-Net model, the pre-trained BEATs model plays a significant role in deploying informative acoustic features. However, due to its model structure, BEATs require computationally intensive resources to perform feature extraction. This dependency introduces significant computational overhead, especially in scenarios that require real-time processing or low-latency inference.
To this end, our focus shifted to developing a more compact and efficient version of HAAQI-Net, but still ensuring satisfactory performance in music audio quality assessment for hearing aid users.

Knowledge distillation \cite{10.1007/s11263-021-01453-z} emerges as a pivotal technique in our endeavor to transfer expertise from a large, cumbersome model (teacher) to a learner, more lightweight counterpart (student). Our goal is to distill the rich representations acquired by the BEATs model in the teacher network into the student model while ensuring that performance is not affected. 
To achieve this goal, as shown in Fig.~\ref{fig:distillHAAQI}, we perform a multi-stage distillation process beyond the final output layer to cover multiple intermediate layers of the teacher model. This strategic approach ensures that the student model fully grasps the teacher model's ability to extract features and potentially maintains generalization. 

Furthermore, we leverage the power of transfer learning \cite{9134370} to initialize the weights of the student model using pre-trained parameters of the corresponding layers of the teacher model. This strategic integration allows the student model to exploit the knowledge encoded in the teacher model parameters, resulting in faster convergence of the learning process. In addition, as shown in Fig.~\ref{fig:distillHAAQI}, we retain fundamental modules such as pre-processing, patch embedding, layer normalization, and a subset of Transformer encoder layers while discarding redundant elements. In this way, we assume that effective knowledge distillation can be achieved, potentially optimizing performance in music audio quality assessment tasks for hearing aid users. In the following part, we will introduce our adaptive distillation strategy.

\subsubsection{Adaptive Distillation}
Adaptive distillation is beneficial in situations where the complexity or difficulty of samples in the training data varies significantly. In traditional distillation methods, the student model learns from a fixed teacher model, and all training samples are treated equally in terms of their contribution to the loss function. However, this may not be optimal when dealing with diverse datasets, where some samples are more difficult than others. Adaptive distillation solves this problem by dynamically adjusting the loss weight based on the estimated difficulty of each training sample.

In our model, the adaptive distillation loss function works by first calculating a difficulty measure for each training sample, typically based on a similarity measure between the student model's prediction and the ground truth label, such as cosine similarity. Training samples with a higher similarity measure are considered easier, while those with a lower similarity measure are deemed more difficult. The loss for each training sample is then weighted according to its difficulty measure. More difficult training samples receive higher weights in the loss function, allowing the model to focus more on learning from challenging training samples. Finally, the weighted losses are aggregated to calculate the distillation loss for the batch:
\begin{align}
\label{eq:Distil_loss}
    L_{Distil}  = \frac{1}{B} \sum_{n=1}^{B} \left( \frac{1}{3} \sum_{i=6, 9, 12} L_{TE\_i,n} \right) \times d_n,
\end{align}
where $d_{n}$ is the difficulty weight for training sample $n$, and $L_{TE\_i,n}$ is the layer-wise distillation loss for the $i$-th layer of training sample $n$ calculated as
\begin{align}
\label{eq:L_TE_i}
    L_{TE\_i,n}  =  L_{L1,n}^{i} + L_{cos,n}^{i}.
\end{align}
The layer-wise L1 loss $L_{L1,n}^{i}$ is calculated as
\begin{align}
\label{eq:L_L1}
    L_{L1,n}^{i}  =  \left| \text{\textbf{X}}_{\text{BEATs},n}^i - \text{\textbf{X}}_{\text{TE $i$ Predictor},n}\right|,
\end{align}
where $\textbf{X}_{\text{BEATs},n}^i$ and $\textbf{X}_{\text{TE $i$ Predictor},n}$ are the output features from the $i$-th layer of BEATs’ Transformer encoder and the TE $i$ Predictor of the distilled model for training sample $n$, respectively. $L_{cos,n}^{i}$ is the layer-wise sigmoid cosine loss for the $i$-th layer of training sample $n$ calculated as, 
\begin{align}
     L_{cos,n}^{i} & = - s_{n,i} \times \log(\hat{s}_{n,i}) - (1 - s_{n,i}) \times \log(1 - \hat{s}_{n,i}),
\end{align}
where $s_{n,i}$ is the cosine similarity between $\textbf{X}_{\text{BEATs},n}^i$ and $\textbf{X}_{\text{TE $i$ Predictor},n}$, 
 
and $\hat{s}_{n,i}$ is its sigmoid transformation calculated as,
\begin{align}
    \hat{s}_{n,i} = \frac{1}{1 + e^{-s_{n,i}}}.
\end{align}

By combining L1 loss and cosine loss in Eq.~(\ref{eq:L_TE_i}), we aim to exploit their complementary properties to capture
different aspects of the difference between predicted and true BEATs features.
The overall loss for training HAAQI-Net with knowledge distillation is as follows:
\begin{align}
     L_{\text{}} = L_{Qual} + L_{Distil}.
\end{align}

\section{Experiments}
\label{sec:Numerical Experiments}
This section details the experiments performed in this work, covering aspects such as data preparation, experimental setup, and obtained results.

\subsection{Data preparation}
Before delving into the evaluation of HAAQI-Net, we first explain how music samples are (i) selected, (ii) processed, and  (iii) how hearing loss patterns are generated. 

\subsubsection{Music Samples}
The music dataset is based on the small split of the FMA (FMA-small) dataset \cite{Defferrard2016FMAAD} and the MTG Jamendo dataset \cite{bogdanov2019mtg}. FMA-small is a balanced dataset for genre classification. From the eight genres available in FMA-small, we selected five genres: hip-hop, instrumental, international, pop, and rock. Considering that people with hearing loss are more likely to be older adults who regularly listen to classical music and orchestral music \cite{bonneville2013music}, we added additional music samples of these two genres from the MTG-Jamendo dataset.
Through random selection, we collected 600 mono-audio samples in 7 genres, each lasting 30 seconds.

\subsubsection{Music Signal Processing}
The HAAQI-Net model is trained using data processed according to the method outlined in \cite{arehart2010effects}. Their study aimed to explore how various signal processing techniques, typical in commercial hearing aids, affect judgments of music quality. Their experiments involved 100 distinct processing conditions in three groups:

\begin{itemize}[]
    \item $32$ conditions involving noise addition and nonlinear processing under different SNRs and parameter settings.
    \item $32$ conditions employing linear filtering under different settings.
    \item $36$ conditions combining noise addition, nonlinear processing, and linear filtering under different settings.
\end{itemize}

Noise conditions included music embedded in stationary speech-shaped noise and multi-talker babble. Nonlinear processing encompassed symmetric peak clipping, amplitude quantization, Wide Dynamic Range Compression (WDRC), spectral subtraction noise suppression, and various combinations. Linear processing involved low-pass and high-pass filters, spectral tilt adjustment, resonance peaks, and combinations of band-pass filters with resonance peaks. 
The combined processing conditions included all possible combinations of the above conditions.

\subsubsection{Hearing Loss Patterns}
\begin{figure}[t]
  \centering
  \resizebox{\linewidth}{!}{\input{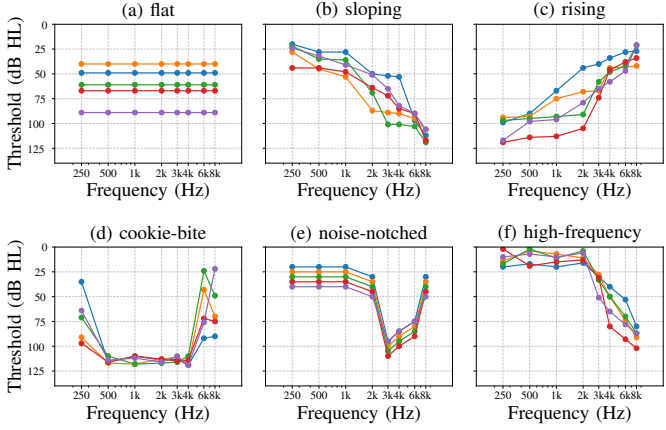}}
  \caption{Some examples of hearing loss audiograms: the $y$-axis represents the hearing threshold in dB, and the $x$-axis represents the frequency in Hz.}
  \label{fig:res}
\end{figure}

When generating hearing loss patterns, we followed the approach in \cite{alshuaib2015classification}.
Hearing loss is described through an audiogram, which is a graphical display showing the degrees of hearing loss in different frequency regions, as shown in Fig.~\ref{fig:res}.
A threshold above $20$dB at any frequency is considered hearing loss. 
We represent a specific pattern of hearing loss using an $8$-dimensional vector, where each dimension represents the threshold at that specific frequency.
The eight frequencies used are $250$, $500$, $1$K, $2$K, $3$K, $4$K, $6$K, and $8$KHz.
There are $6$ types of audiograms: flat, sloping, rising, cookie-bite, noise-notched, and high-frequency, as illustrated in Fig.~\ref{fig:res}.
In total, we established $300$ hearing loss patterns, $50$ patterns per hearing loss category. The patterns of each category were divided into two groups, $40$ patterns for training and the remaining $10$ patterns for testing.

\begin{table}[t]
\centering
\caption{
Statistics of the data used to evaluate HAAQI-Net. 
}
\begin{tabular}{lrrr}
\toprule
\multicolumn{1}{c}{Genre} & \multicolumn{1}{c}{Training Data} & \multicolumn{2}{c}{Testing Data} \\
\cmidrule(lr){3-4}
\multicolumn{1}{c}{} & \multicolumn{1}{c}{} & \multicolumn{1}{c}{Seen Noise} & \multicolumn{1}{c}{Unseen Noise} \\
\midrule
\midrule
\multicolumn{4}{c}{FMA-small} \\
\midrule
Hip-hop & 3,048 & 552 & 116 \\
\midrule
Instrumental & 3,890 & 538 & 104 \\
\midrule
International & 3,368 & 484 & 104 \\
\midrule
Pop & 3,632 & 496 & 110 \\
\midrule
Rock & 4,644 & 396 & 74 \\
\midrule
\midrule
\multicolumn{4}{c}{MTG-Jamendo} \\
\midrule
Classical & 3,830 & 206 & 82 \\
\midrule
Orchestral & 3,388 & 768 & 130 \\
\midrule
\midrule
\textbf{Total} & \textbf{25,800} & \textbf{3,440} & \textbf{720} \\
\bottomrule
\end{tabular}

\label{tab:music_data}
\end{table}

\begin{figure}[t]
  \centering
  \resizebox{\linewidth}{!}{\input{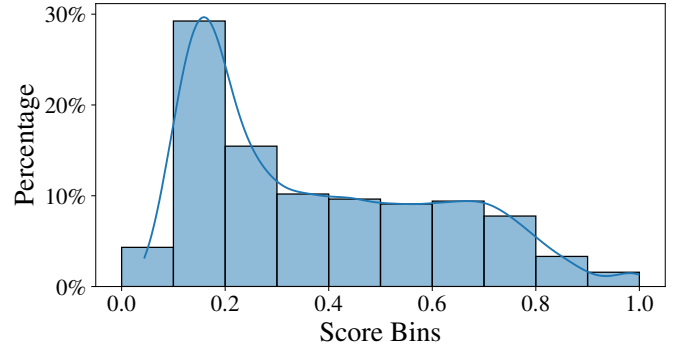}}
  \caption{Distribution of HAAQI scores for the music samples used to evaluate HAAQI-Net.}
  \label{fig:score_distribution}
\end{figure}

Table~\ref{tab:music_data} summarizes the data used to evaluate HAAQI-Net. 
Each music sample was paired with a randomly selected pattern of hearing loss. The test set was divided into seen and unseen subsets, where ``seen'' means the processing type was seen in training. Note that, as mentioned above, the hearing loss patterns in testing are always unseen in training.
The distribution of HAAQI scores for all music samples across different data processing conditions and patterns of hearing loss is shown in Fig. \ref{fig:score_distribution}.

\subsection{Experimental Setup}
\label{sec:Experimental Setup}
In this section, we introduce the experimental setup, including input configurations and various experimental scenarios.

\subsubsection{Inputs and Configurations}
In our experiments, we explored different settings to identify the most effective features for music quality prediction. These settings include spectrogram, speech SSL models, and BEATs features. For the spectrogram features, the power spectral features are extracted via a $512$-point Short-Time Fourier transform (STFT) with a Hamming window size of $512$ points and a hop size of $256$ points, resulting in a $257$-dimensional magnitude spectrum. The sampling rate for this input is $32$ kHz, which is then resampled to $16$ kHz with separate mono channels (left and right) to match the input sampling rate of the BEATs model. This resampling step ensures compatibility with the model while preserving the relevant acoustic features for the task. For speech SSL models, we selected three well-known models: Wav2Vec 2.0 Large \cite{Baevski2020wav2vec2A}, HuBERT Large \cite{Hsu2021HuBERTSS}, and WavLM Large \cite{Chen2021WavLMLS}. Specifically, we used the last layer of these models as the acoustic features. As for the BEATs \cite{BEATs} features, we used four different configurations: ``BEATs (Last)'', ``BEATs (Last) + Win Avg'', ``BEATs (Last) + Adapter'', and ``BEATs (WS) + Adapter''. In detail, ``BEATs (Last)'' uses the last layer of the Transformer encoder as acoustic features with dimension $768$. ``BEATs (Last) + Win Avg'' applies a moving average for every three elements so that the feature dimension is comparable to that of the spectrogram features. ``BEATs (Last) + Adapter'' employs an adapter (i.e., a dense layer) to reduce the dimensionality of the BEATs features (the last layer output) to $257$. The adapter also plays a role in adapting the BEATs features for music audio quality assessment tasks. ``BEATs (WS) + Adapter'' employs an adapter to reduce the dimensionality of the BEATs features (the weighted sum of all layers) to $257$. Notably, many hearing aids operate at a 16 kHz sampling rate, as discussed on \cite{Kates2008}. This is a common sampling rate used in hearing aid devices, which also typically represent signal samples with 16-bit words. Given this context, we chose the 16 kHz sampling rate to align with practical constraints in hearing aid technology and to ensure our model's evaluation is relevant to real-world applications.

Each type of acoustic input is concatenated with the hearing-loss pattern to form the final input to HAAQI-Net. The corresponding ground-truth quality score is calculated by the HAAQI method, ranging from $0$ to $1$, with $0$ indicating poor quality and $1$ representing perfect quality. The stimuli were amplified using the National Acoustics Laboratories revised (NAL-R) \cite{byrne1986national} linear fitting prescriptive formula based on individual hearing loss patterns.
We trained HAAQI-Net using the Adam optimizer with a learning rate of $10^{-4}$ and an early stopping technique. To evaluate the performance, three criteria are used: LCC, SRCC, and MSE.

\begin{table*}[ht] 
  \centering
  \caption{Performance of music quality prediction of HAAQI-Net using different input features.}  
  \label{tab:hearing_loss_performance1}
  \begin{adjustbox}{width=\textwidth} 
    \small 
    \setlength{\tabcolsep}{4pt} 
    \begin{tabularx}{\linewidth}{l*{9}{>{\centering\arraybackslash}X}}
      \toprule
      \multirow{2}{*}{Input Features} & \multicolumn{3}{c}{All} & \multicolumn{3}{c}{Seen} & \multicolumn{3}{c}{Unseen} \\
      \cmidrule(lr){2-4} \cmidrule(lr){5-7} \cmidrule(lr){8-10}
      & LCC $\uparrow$ & SRCC $\uparrow$ & MSE $\downarrow$ & LCC $\uparrow$ & SRCC $\uparrow$ & MSE $\downarrow$ & LCC $\uparrow$ & SRCC $\uparrow$ & MSE $\downarrow$ \\
      \midrule
      \midrule
      Spectrogram               & 0.6701 & 0.7084 & 0.0301 & 0.6848 & 0.7106 & 0.0298 & 0.5999 & 0.6612 & 0.0314 \\
      \midrule
      Wav2Vec 2.0 Large (Last) & 0.6483 & 0.6643 & 0.0302 & 0.6651 & 0.6809 & 0.0304 & 0.5030 & 0.5268 & 0.0295 \\
      \midrule
      HuBERT Large (Last)      & 0.6392 & 0.6645 & 0.0336 & 0.6396 & 0.6642 & 0.0343 & 0.5717 & 0.5478 & 0.0305 \\
      \midrule
      WavLM Large (Last)       & 0.7809 & 0.7834 & 0.0202 & 0.7806 & 0.7862 & 0.0214 & 0.7050 & 0.6757 & 0.0145 \\
      \midrule
      BEATs (Last)              & 0.9274 & 0.9342 & 0.0071 & 0.9234 & 0.9332 & 0.0081 & 0.9014 & 0.8762 & 0.0026 \\
      \midrule
      BEATs (Last) + Win Avg  & 0.8765 & 0.8908 & 0.0123 & 0.8724 & 0.8879 & 0.0135 & 0.8119 & 0.8275 & 0.0061 \\
      \midrule
      BEATs (Last) + Adapter     & 0.9368 & 0.9486 & 0.0064 & 0.9327 & 0.9455 & 0.0073 & 0.9282 & 0.9188 & 0.0024 \\
      \midrule
      \textbf{BEATs (WS) + Adapter*} & \textbf{0.9456} & \textbf{0.9603} & \textbf{0.0055} & \textbf{0.9410} & \textbf{0.9568} & \textbf{0.0063}  & \textbf{0.9518} & \textbf{0.9417} & \textbf{0.0014}\\
      \bottomrule
    \end{tabularx}
  \end{adjustbox}
  \vspace{5pt} 
  \par\noindent\footnotesize{\textit{Note:} The arrows ($\uparrow$/$\downarrow$) indicate the direction of better performance for each metric: higher values are better for LCC and SRCC, while lower values are better for MSE. *All differences between BEATs (WS) + Adapter and other models are statistically significant based on a t-test ($p \textless 0.05$).}
\end{table*}

\subsubsection{Experimental Scenarios}
We explored two scenarios to evaluate the generalization ability of HAAQI-Net: the seen set and the unseen set. The seen set contains data with the same music processing conditions as the training set, while the unseen set contains data with different music processing conditions than the training set. We selected $82$ conditions for the seen set, leaving $18$ conditions exclusive for the unseen set. The unseen set comprises conditions such as ``compression + babble'', ``compression + spectral subtraction + babble'', ``multiple resonance peaks + low pass filter'', ``babble + compression + high pass filter'', ``babble + compression + low pass filter'', ``babble + compression + positive spectral tilt'', ``babble + compression + negative spectral tilt'', ``babble + compression + single resonance peak'', and ``babble + compression + multi-resonance peak''. The distribution of training and testing data for each genre, as well as the seen and unseen sets, are summarized in Table \ref{tab:music_data}. $80\%$ of the training data is used for training and $20\%$ for validation.

\subsection{Experimental Results}
For a thorough grasp of the model's performance, we analyze (i) overall performance, (ii) scenario-based performance, and (iii) efficiency.

\subsubsection{Overall Performance with Different Input Features} 
Table \ref{tab:hearing_loss_performance1} shows the music quality prediction performance of HAAQI-Net using different input features. We first focus on the overall performance (see ``All'' in the table) and make the following observations. First, the spectrogram  provides a standard representation of audio signals but lacks context, resulting in moderate performance. Second, Wav2ec 2.0 and HuBERT, although effective in many speech processing tasks, show inferior performance, possibly due to their general-purpose nature. Third, WavLM, designed for automatic speech recognition, outperforms other speech SSL models by capturing higher-level semantic information. Fourth, BEATs trained with different audio data outperforms spectrogram and all three speech SSL models. Fifth, dimension reduction based on moving average will reduce the performance (``BEATs (Last) + Win Avg'' vs ``BEATs (Last)''), but the low-dimensional features are still more effective than the spectrogram (``BEATs (Last) + Win Avg'' vs spectrogram). Sixth, a simple adapter can effectively adapt the BEATs features for music quality prediction (``BEATs (Last) + Adapter'' vs ``BEATs (Last)'').
Seventh, ``BEATs (WS) + Adapter'' is the most effective among all input feature configurations compared here.

\begin{figure}[t]
    \centering
    \includegraphics[width=\linewidth]{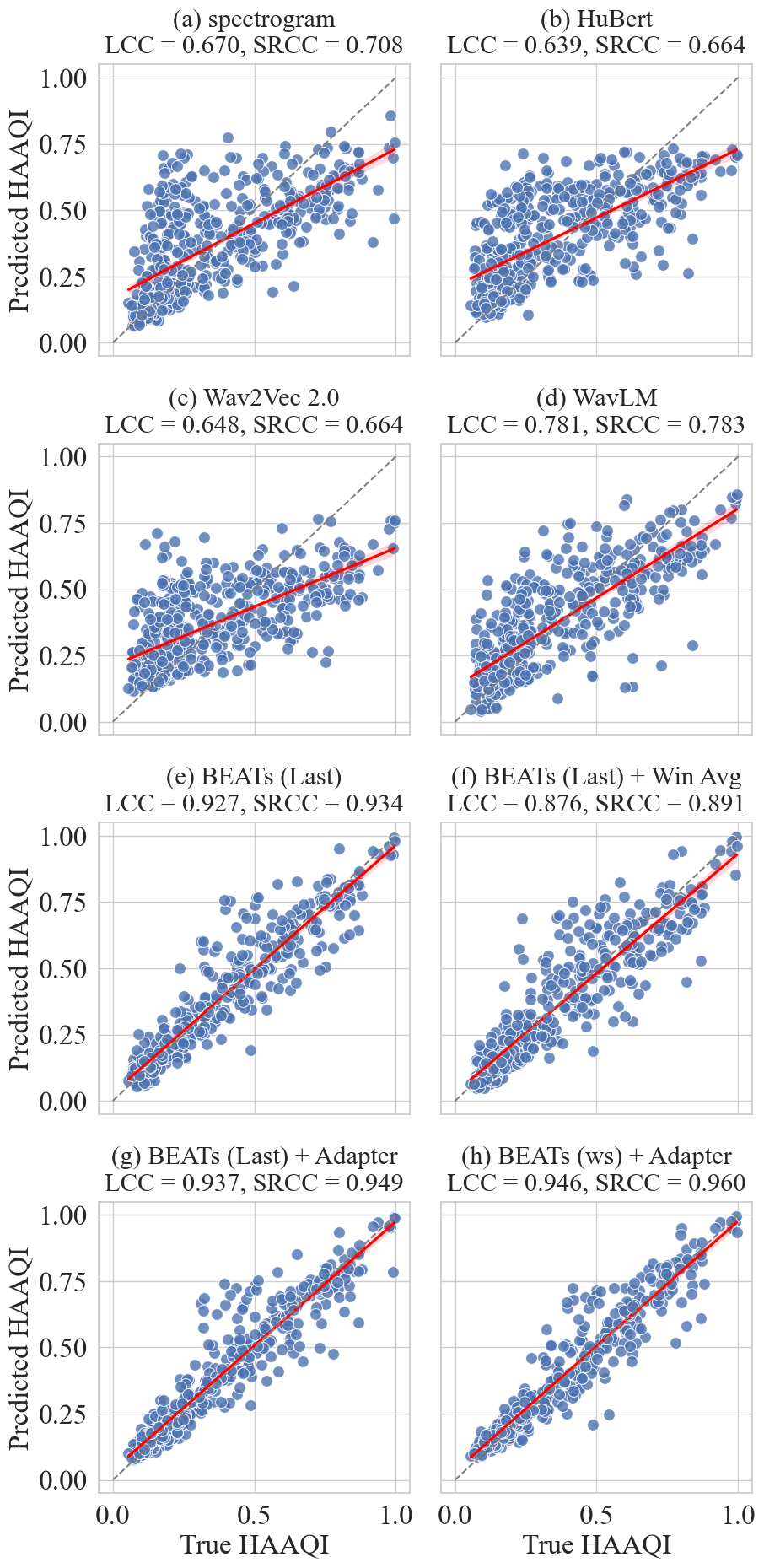}
\caption{Scatter plots of music quality prediction of HAAQI-Net using different input features. The dashed diagonal line represents the optimal prediction, while the red line represents the regression line with 95\% confidence interval for the model predictions. Data points below the dashed diagonal line indicate that the model's predictions are lower than the true HAAQI scores, while data points above the line indicate that the model's predictions are higher.}
  \label{fig:haaqi_net_res}
\end{figure}

As shown in the scatter plots in Fig. \ref{fig:haaqi_net_res}, it is clear that among all the features compared here, the predictions of HAAQI-Net using the ``BEATs (WS) + Adapter'' features are most concentrated near the optimal diagonal. In addition, it can be seen that the performance of various BEATs features is better than that of spectrogram and three speech SSL model features.

\begin{table*}[ht]
  \centering
  \caption{Performance of HAAQI-Net under different types of hearing loss.}
  \label{tab:hearing_loss_performance2}
  \adjustbox{max width=\textwidth}{
    \small
    \begin{tabular}{l *{13}{r}}
      \toprule
      \multirow{2}{*}{Hearing loss type} & \multicolumn{3}{c}{Spectrogram} & \multicolumn{3}{c}{BEATs (Last)} & \multicolumn{3}{c}{BEATs (Last) + Adapter} & \multicolumn{3}{c}{BEATs (WS) + Adapter} \\
      \cmidrule(lr){2-4} \cmidrule(lr){5-7} \cmidrule(lr){8-10} \cmidrule(lr){11-13}
      & LCC $\uparrow$ & SRCC $\uparrow$ & MSE $\downarrow$ & LCC $\uparrow$ & SRCC $\uparrow$ & MSE $\downarrow$ & LCC $\uparrow$ & SRCC $\uparrow$ & MSE $\downarrow$ & LCC $\uparrow$ & SRCC $\uparrow$ & MSE $\downarrow$ \\
      \midrule
      \midrule
      Flat              & 0.6129 & 0.6445 & 0.0407 & 0.9320 & 0.9387 & 0.0074 & 0.9503 & 0.9552 & 0.0054 & \textbf{0.9589} & \textbf{0.9630} & \textbf{0.0047} \\
      \midrule
      Sloping           & 0.6520 & 0.6919 & 0.0332 & 0.9263 & 0.9373 & 0.0072 & 0.9234 & 0.9424 & 0.0080 & \textbf{0.9411} & \textbf{0.9593} & \textbf{0.0060} \\
      \midrule
      Rising            & 0.7104 & 0.7432 & 0.0175 & 0.9174 & 0.9248 & 0.0056 & 0.9279 & 0.9501 & 0.0050 & \textbf{0.9497} & \textbf{0.9608} & \textbf{0.0035} \\
      \midrule
      Cookie-bite       & 0.7192 & 0.7401 & 0.0111 & 0.8101 & 0.8169 & 0.0058 & \textbf{0.8791} & 0.8547 & \textbf{0.0039} & 0.8757 & \textbf{0.9007} & 0.0040 \\
      \midrule
      Noise-notched     & 0.6439 & 0.6726 & 0.0278 & 0.9350 & 0.9479 & 0.0057 & 0.9326 & 0.9494 & 0.0064 & \textbf{0.9413} & \textbf{0.9568} & \textbf{0.0050} \\
      \midrule
      High-frequency    & 0.5864 & 0.5862 & 0.0400 & 0.9149 & 0.9263 & 0.0090 & 0.9401 & 0.9432 & \textbf{0.0068} & \textbf{0.9412} & \textbf{0.9584} & 0.0069 \\
      \bottomrule
    \end{tabular}
  }
\end{table*}

\subsubsection{Scenario-Based Performance}
To fully evaluate the generalization ability of HAAQI-Net, we examine its performance on the seen and unseen test sets (see ``Seen'' and ``Unseen'' in Table \ref{tab:hearing_loss_performance1}). 
From the table, we can see that under different input feature configurations, HAAQI-Net generally performs worse on the unseen test set than on the seen test set as expected. Various BEATs features outperform the spectrogram and three SSL model features on both seen and unseen test sets. The adapter for adapting the BEATs features for music quality prediction appears to be effective in reducing the performance gap between the seen and unseen test sets. Surprisingly, HAAQI-Net with ``BEATs (WS) + Adapter'' performs even slightly better on the unseen test set than the seen test set in terms of MSE and LCC. Again, ``BEATs (WS) + Adapter'' is the most effective among all input feature configurations compared here.

\subsubsection{Performance under Different Hearing Loss Patterns}
The performance of HAAQI-Net with different input features under different hearing loss patterns is shown in Table \ref{tab:hearing_loss_performance2}. We can see that the ``BEATs (Last)'' features are more effective than the spectrogram across all hearing loss types. The adapter for adapting the ``BEATs (Last)'' features for music quality prediction appears to be effective at boosting the performance (``BEATs (Last) + Adapter'' vs ``BEATs (Last)''). The weighted-sum approach provides more informative features than using the last layer directly, as ``BEATs (WS) + Adapter'' outperforms ``BEATs (Last) + Adapter'' for most types of hearing loss. It is worth noting that HAAQI-NET performs slightly worse under the cookie-bite loss type than other loss types.

\begin{figure*}[ht]
  \centering
  \subfloat{\resizebox{0.5\textwidth}{!}{\input{image/pgfplot/result_all_test_ws_noise_1_2.pgf}}}
  \hfill
  \subfloat{\resizebox{0.5\textwidth}{!}{\input{image/pgfplot/result_all_test_ws_noise_3_8.pgf}}}
  \caption{Performance of HAAQI-Net with ``BEATs (WS) + Adapter'' under different signal processing conditions. Each data point in the plot represents a specific signal processing condition, averaged across all combinations of music samples and hearing loss patterns. The horizontal and vertical error bars represent the standard deviation of the true HAAQI scores and the predicted HAAQI scores, respectively. In (c), LTASS* represents the combination of LTASS with 6 of the linear filtering methods in (b), and similarly for (d)-(h).}
  \label{fig:HAAQI_on_processing}
\end{figure*}


\subsubsection{Performance under Different Signal Processing Conditions}
Fig. \ref{fig:HAAQI_on_processing} shows the performance of HAAQI-Net with the ``BEATs (WS) + Adapter'' features under different signal processing conditions.

The 8 processing conditions in Fig. \ref{fig:HAAQI_on_processing}(a) correspond to the noise addition and non-linear processing in \cite{arehart2010effects}.
For each condition, the result is an average of different settings, such as different SNRs and clipping thresholds.
``LTASS'' represents music with additive stationary speech-shaped noise, ``Babble'' denotes music with multi-talker babble, ``Peak Clip'' indicates music subjected to symmetric instantaneous peak clipping, ``Quant'' represents music quantized with reduced bit depth, ``Comp'' refers to music processed through multi-channel WDRC, ``Comp+Babble'' signifies music processed through WDRC after adding babble, ``SSub+Babble'' denotes music processed by spectral subtraction after adding babble, and ``Comp+SS+Babble'' refers to music processed by WDRC and spectral subtraction after adding babble.
In Fig. \ref{fig:HAAQI_on_processing}(a), 
the ``Comp'' condition shows relatively high true HAAQI mean and standard deviation values, while its predicted HAAQI values are closely aligned. This suggests that the model performs consistently and accurately under the ``Comp'' condition. The results under the ``Quant'' condition are similar to those under the ``Comp'' condition. 
The results under the ``LTASS'' and ``Peak Clip'' conditions are close to each other, but the HAAQI scores are lower compared to the ``Comp'' and ``Quant'' conditions. 
In contrast, conditions including ``Babble'', ``Comp+Babble'', ``SSub+Babble'' and ``Comp+SS+Babble'' show lower true and predicted HAAQI means, as well as lower standard deviations, indicating that the corresponding signal processing methods consistently have a large impact on music quality. Overall, the results highlight the varying HAAQI scores (music quality) under noise addition and non-linear processing, and the close alignment between predicted and true HAAQI values demonstrates the superior performance of HAAQI-Net.

The 8 processing conditions in Fig. \ref{fig:HAAQI_on_processing}(b) pertain to the linear filtering conditions in \cite{arehart2010effects}. 
For each condition, the result is an average of different settings, such as different pass bands.
``HP Filt'' represents music processed by a high-pass filter, ``LP Filt'' denotes music processed by a low-pass filter, ``BP Filt'' represents music processed by a band-pass filter, ``Pos Tilt'' represents music processed through a filter with positive spectral tilt, ``Neg Tilt'' denotes music processed through a filter with negative spectral tilt, ``Single Peak'' indicates music processed through a filter with a single spectral peak, ``Multi Peak'' represents music processed through a filter with three spectral peaks, and ``Multi Peak+LP Filt'' represents music processed sequentially through a filter with three spectral peaks and a low-pass filter.
From Fig.~\ref{fig:HAAQI_on_processing}(b), we can also see some similar trends to Fig.~\ref{fig:HAAQI_on_processing}(a). For example, different filtering conditions can result in significantly different HAAQI scores. Conditions with higher HAAQI means also show higher standard deviations. Except for the ``LP Filt'', ``HP Filt'', and ``Single Peak'' conditions, all other more complex filtering conditions result in lower HAAQI scores. Overall, except for the ``HP Filt'' condition, HAAQI-Net predictions are in good agreement with the true HAAQI scores. 


The 36 processing conditions in Fig. \ref{fig:HAAQI_on_processing}(c-h) correspond to possible combinations of 6 ``noise addition and nonlinear
processing'' methods and 6 ``linear filtering'' methods, as settings in \cite{arehart2010effects}. Each plot shows the results of a subset that combines a specific ``noise addition and nonlinear
processing'' method with the 6 ``linear filtering'' methods shown in the legend. Although some data points appear to be slightly off the diagonal, such as those corresponding to the combination of one of ``noise addition and nonlinear processing'' methods with ``Pos Tilt'' (e.g., the combination of ``Quant'' and ``Pos Tilt'' in Fig. \ref{fig:HAAQI_on_processing}(f)), the performance of HAAQI-Net is generally good.   


The noise and nonlinear conditions for the combined data mirror those of the noise and distortion data. In the "Comp+Babble" category, for instance, true and predicted HAAQI means are relatively close, with slight variations observed across different sub-conditions such as "Comp+HP Filt", "Comp+LP Filt", "Comp+Multi Peak", "Comp+Neg Tilt", "Comp+Pos Tilt", and "Comp+Single Peak". Notably, "Comp+Babble" sub-conditions generally exhibit similar patterns, with true and predicted means closely aligned. Similarly, in the "Babble" category, true and predicted HAAQI means are comparable across different sub-conditions such as "Babble+HP Filt", "Babble+LP Filt", "Babble+Multi Peak", "Babble+Neg Tilt", "Babble+Pos Tilt", and "Babble+Single Peak", suggesting consistent modeling across these scenarios. Moreover, in the "Peak Clip" and "Quant" categories, true and predicted means exhibit close alignment, albeit with slight deviations observed in certain sub-conditions. However, in the "LTASS" category, while true and predicted means are relatively close, some sub-conditions such as "LTASS+Pos Tilt" show slightly higher deviations. Overall, the analysis suggests that the model effectively predicts HAAQI scores across various processing conditions, with a generally close alignment observed between true and predicted means. This consistency in performance underscores the model's robustness and reliability across different scenarios.

\begin{figure}[t]
  \centering
  \includegraphics[width=\linewidth]{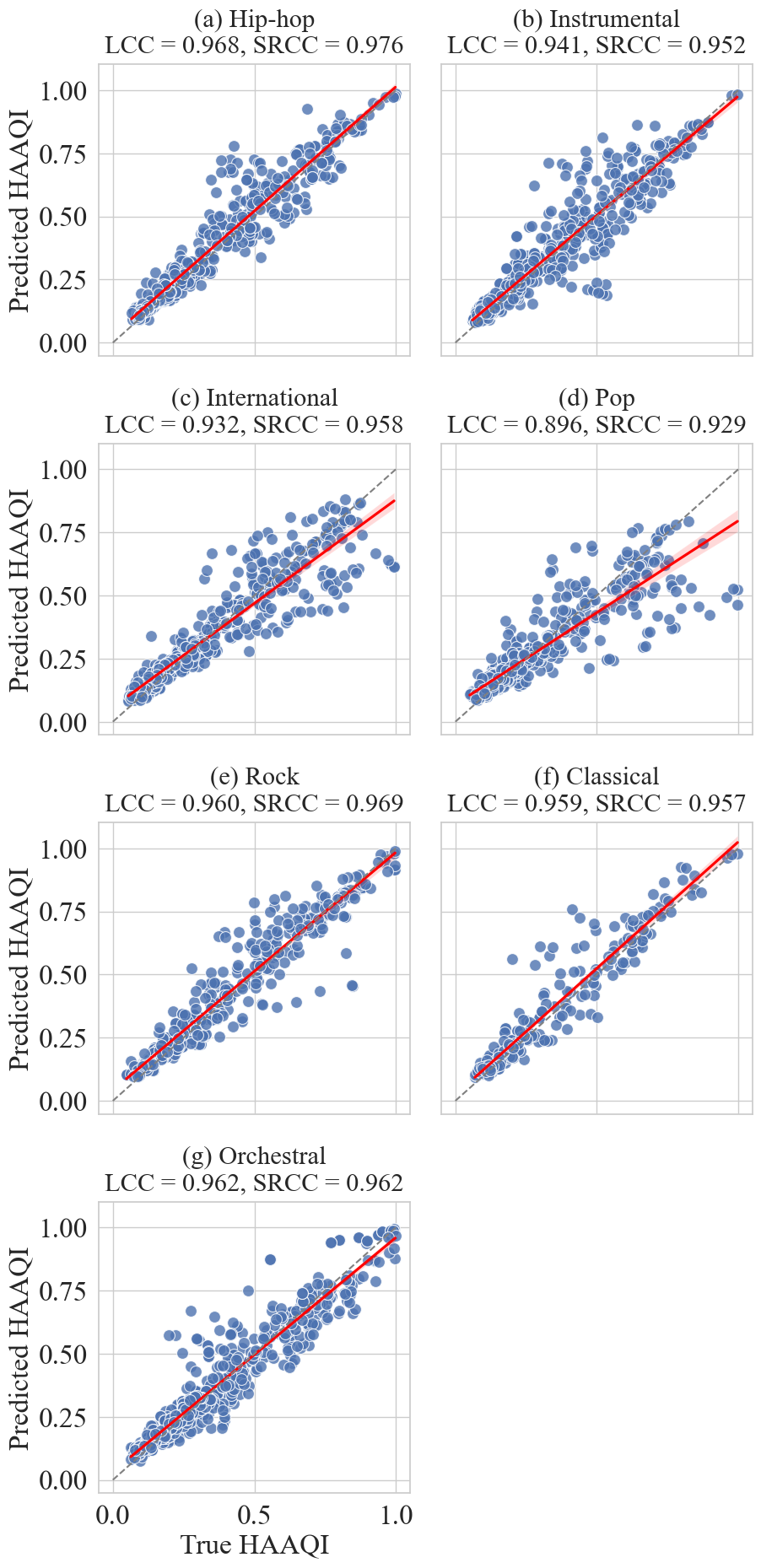}
  \caption{Scatter plots of true HAAQI scores and HAAQI-Net's predicted HAAQI scores for different music genres. The dashed diagonal line represents the optimal prediction, while the red line represents the regression line with 95\% confidence interval for the model predictions.}
  \label{fig:HAAQI_on_genre}
\end{figure}

\subsubsection{Performance across Different Genres} 
Figure \ref{fig:HAAQI_on_genre} shows the scatter plots of music quality prediction of our best-performing model (HAAQI-Net with the ``BEATs (WS) + Adapter'') across various music genres. We can see that the model performs well in all music genres except pop music. In fact, for hip-hop, instrumental, international, pop, rock, classical, and orchestral music, the LCC values are 0.9678, 0.9407, 0.9321, 0.8960, 0.9603, 0.9593, and 0.9615, respectively.
The model's relatively poor performance in pop music may be due to its complex arrangements and electronic instrumentation. For international music, its diverse cultural styles may also pose slight challenges to the model. Overall, the experimental results show that the model has certain generalization capabilities across genres.

\begin{table*}[ht]
\centering
\caption{Performance of HAAQI-Net with Different Distillation Strategies.}
\label{tab:haaqi-net-distillation}
\begin{adjustbox}{width=\textwidth} 
    \small 
    \setlength{\tabcolsep}{4pt} 
    \begin{tabularx}{\linewidth}{p{3.6cm} p{1.7cm}*{9}{>{\centering\arraybackslash}X}}
        \toprule
        \multirow{2}{*}{Distillation Methods} & \multirow{2}{*}{\centering \makecell{Distillation \\ Loss}} & \multicolumn{3}{c}{All} & \multicolumn{3}{c}{Seen} & \multicolumn{3}{c}{Unseen} \\
        \cmidrule(lr){3-5} \cmidrule(lr){6-8} \cmidrule(lr){9-11}
        & & LCC $\uparrow$ & SRCC $\uparrow$ & MSE $\downarrow$ & LCC $\uparrow$ & SRCC $\uparrow$ & MSE $\downarrow$ & LCC $\uparrow$ & SRCC $\uparrow$ & MSE $\downarrow$ \\
        \midrule
        \midrule
        Original HAAQI-Net      & - & 0.9368   & 0.9487   & 0.0064  & 0.9327   & 0.9455   & 0.0073   & 0.9283   & 0.9188   & 0.0024  \\
        \midrule
        \midrule
        Single-layer distillation  & L1 + Cosine  & 0.8043   & 0.8393   & 0.1877  & 0.7907   & 0.8386   & 0.0217   & 0.7619   & 0.7697   & 0.0047  \\
        \midrule
        \midrule
        \multicolumn{11}{c}{Sequential Prediction Heads} \\
        \midrule
        Multi-layer distillation & L1 + Cosine & 0.8904 & 0.9087 & 0.0106 & 0.8865 & 0.9080 & 0.0117 & 0.8729 & 0.8514 & 0.0052 \\
        \midrule
        Multi-layer distillation (fine-tuning BLSTM\&ATT) & L1 + Cosine & 0.8750 & 0.9018 & 0.0135 & 0.8694 & 0.8938 & 0.0147 & 0.8724 & 0.8728 & 0.0081 \\
        \midrule
        Multi-layer distillation & Adaptive & 0.8963 & 0.9155 & 0.0108 & 0.8923 & 0.9107 & 0.0118 & 0.8704 & 0.8802 & 0.0063 \\
        \midrule
        \midrule
        \multicolumn{11}{c}{Independent Prediction Heads} \\
        \midrule
        Multi-layer distillation & L1 + Cosine & 0.8988 & 0.9105 & 0.0098 & 0.8919 & 0.9089 & 0.0114 & 0.9054 & 0.8876 & 0.0024 \\
        \midrule
        Multi-layer distillation (fine-tuning BLSTM\&ATT) & L1 + Cosine & 0.8951 & 0.9217 & 0.0103 & 0.8898 & 0.9149 & 0.0115 & 0.8961 & 0.8927 & 0.0048 \\
        \midrule
        Multi-layer distillation & Adaptive & 0.9071 & 0.9307 & 0.0091 & 0.8997 & 0.9250 & 0.0106 & 0.9116 & 0.8994 & 0.0019 \\
        \midrule
        Multi-layer distillation with ws & Adaptive & 0.9151 & 0.9331 & 0.0083 & 0.9083 & 0.9281 & 0.0096 & 0.9127 & 0.8980 & 0.0022 \\
        \bottomrule
    \end{tabularx}
\end{adjustbox}
\end{table*}

\subsubsection{Performance of HAAQI-Net with Knowledge Distillation}
We conduct an extensive analysis of the performance of HAAQI-Net under various distillation strategies.
The results are shown in Table~\ref{tab:haaqi-net-distillation}. 

The original HAAQI-Net, whose predictions exhibit high correlation and low MSE value with true HAAQI scores in above experiments, serves as the benchmark. Two loss functions are studied: ``L1 + Cosine'' loss and ``Adaptive'' loss. The former combines ``L1'' loss and ``Cosine'' loss, aiming to exploit their complementary properties to capture different aspects of the difference between predicted and true BEATs features. It is calculated using Eq.~(\ref{eq:Distil_loss}) without the difficulty weights of the training samples. In contrast, the ``Adaptive'' loss is directly calculated by Eq.~(\ref{eq:Distil_loss}). It dynamically adapts to the difficulty of each training sample during training and has unique advantages in scenarios with varied sample complexity. In addition, we compare three architectures for distilling the transformer encoder. The first is single-layer distillation that uses only one prediction head (i.e., ``TE 12 Predictor'') connected after ``Transformer Encoder 3'' to imitate the output of ``Transformer Encoder 12'' of the teacher model (see Fig.~\ref{fig:distillHAAQI}). The second is multi-layer distillation with independent prediction heads. As shown in Fig.~\ref{fig:distillHAAQI}, the three independent prediction heads (i.e., ``TE 6 Predictor'', ``TE 9 Predictor'', and ``TE 12 Predictor'') connected after ``Transformer Encoder 3'' respectively simulate the outputs of ``Transformer Encoder 6'', ``Transformer Encoder 9'', and ``Transformer Encoder 12'' of the teacher model. The third is multi-layer distillation with sequential prediction heads, i.e., ``TE 6 Predictor'' imitates the output of ``Transformer Encoder 6'' based on ``Transformer Encoder 3'', ``TE 9 Predictor'' imitates the output of ``Transformer Encoder 9'' based on ``TE 6 Predictor'', and ``TE 12 Predictor'' imitates the output of ``Transformer Encoder 12'' based on ``TE 9 Predictor''. 
We also investigate whether the BLSTM and ATT layers should be fixed or fine-tuned during the distillation process, and whether we should use the weighted sum of the outputs of ``TE 6 Predictor'', ``TE 9 Predictor'', and ``TE 12 Predictor'' or the output of ``TE 12 Predictor''. 


Several observations can be drawn from Table~\ref{tab:haaqi-net-distillation}. 
First, when both use the ``L1 + Cosine'' loss, multi-layer distillation outperforms single-layer distillation regardless of whether sequential prediction heads or independent prediction heads are used. 
Second, fine-tuning BLSTM and ATT layers during the distillation process does not lead to performance gains (see Multi-layer distillation (fine-tuning BLSTM \& ATT) vs. Multi-layer distillation).
Third, multi-layer distillation
with independent prediction heads generally outperforms multi-layer distillation with sequential prediction heads.
Fourth, the ``Adaptive'' loss is more effective the ``L1 + Cosine'' loss (see Multi-layer distillation/Adaptive vs. Multi-layer distillation/L1+Cosine).
Fifth, the weight-sum method is better than simply using the output of “TE 12 Predictor” (see Multi-layer distillation with ws/Adaptive vs. Multi-layer distillation/Adaptive).
Overall, the performance of the distilled HAAQI-Net under the best configuration setting is very close to that of the original HAAQI-Net (see Multi-layer distillation with ws/Adaptive vs. Original HAAQI-Net), while the number
of model parameters is reduced by 75.84\%. 



\subsubsection{HAAQI-Net Tested on the MUSDB18-HQ Dataset}
The MUSDB18-HQ dataset~\cite{MUSDB18HQ} is a widely used dataset in the field of music source separation and quality assessment. It is a high-quality version of the MUSDB18 dataset, which contains professionally produced music tracks with isolated musical sources (e.g., vocals, drums, bass, and other instruments). 
As shown in Table~\ref{tab:musdb18-hq}, the MUSDB18-HQ dataset contains 50 music clips in 9 genres. It is worth noting that most of the music genres in the MUSDB18-HQ dataset are unseen in the training data of the HAAQI-Net models, which adds additional difficulty to the evaluation.
In our experiment, the audio files are corrupted with randomly selected unseen noises. These noises include ambient sounds commonly encountered in urban environments, such as cafe chatter, street noise, bus rumble, and pedestrian footsteps. By incorporating these unseen noises, we aim to simulate real-world scenarios where music audio quality assessment models need to perform robustly in the presence of environmental disturbances.

Table~\ref{tab:musdb18-hq_result} shows the performance of three representative versions of HAAQI-Net tested on the MUSDB18-HQ dataset, including the original HAAQI-Net model (corresponding to  Original HAAQI-Net in Table~\ref{tab:haaqi-net-distillation}) and two reduced HAAQI-Net models, namely HAAQI-Net with distillBEATs$^\dag$ (corresponding to Multi-layer distillation/Adaptive in Table~\ref{tab:haaqi-net-distillation}) and HAAQI-Net with distillBEATs (corresponding to Multi-layer distillation with ws/Adaptive in Table~\ref{tab:haaqi-net-distillation}). Due to unseen genres and noises, we can see that the performance of all three HAAQI-Net models drops when tested on the MUSDB18-HQ dataset (e.g., the LCC of Original HAAQI-Net, HAAQI-Net with distillBEATs$^\dag$, and HAAQI-Net with distillBEATs drops from $0.9368$, $0.9151$, and $0.9071$ to $0.7996$, $0.6059$, and $0.6432$). The results indicate that to be more reliable, HAAQI-Net needs to be trained using training data covering more types of genres and noises. HAAQI-Net with distillBEATs outperforms HAAQI-Net with distillBEATs$^\dag$, this trend is consistent with the results in Table~\ref{tab:haaqi-net-distillation}.
Although the performance gap between HAAQI-Net with distillBEATs and Original HAAQI-Net becomes larger, considering the former's reduced parameter size and important role as a distilled version of a large model, future research can focus on improving the distillation process.

\begin{table}[t]
  \centering
  \caption{Statistics of the MUSDB18-HQ Dataset.}
    \begin{tabular}{lc}
    \toprule
    Genre & Count \\
    \midrule
    \midrule
    Electronic & 10 \\
    Rock & 12 \\
    Folk & 5 \\
    Pop & 4 \\
    Metal & 3 \\
    Hip-Hop & 2 \\
    Blues & 1 \\
    International & 1 \\
    Indie & 11 \\
    \midrule
    \midrule
    Total & 50 \\
    \bottomrule
    \end{tabular}%
  \label{tab:musdb18-hq}%
\end{table}%

\begin{table}[t]
\centering
\caption{Performance of 
HAAQI-Net tested on the MUSDB18-HQ Dataset. 
}
\begin{tabular}{lccc}
\toprule
Models & LCC $\uparrow$ & SRCC $\uparrow$ & MSE $\downarrow$ \\
\midrule
\midrule
Original HAAQI-Net & 0.7996  & 0.7633 & 0.0343 \\
HAAQI-Net with distillBEATs$^\dag$ & 0.6059  & 0.5996 & 0.0482 \\
HAAQI-Net with distillBEATs & 0.6432  & 0.6561 & 0.0507 \\
\bottomrule
\end{tabular}

\label{tab:musdb18-hq_result}
\end{table}

\subsubsection{Adapting HAAQI-Net to Predict Subjective Score}
\begin{figure}[t]
    \centering
    \resizebox{\linewidth}{!}{\input{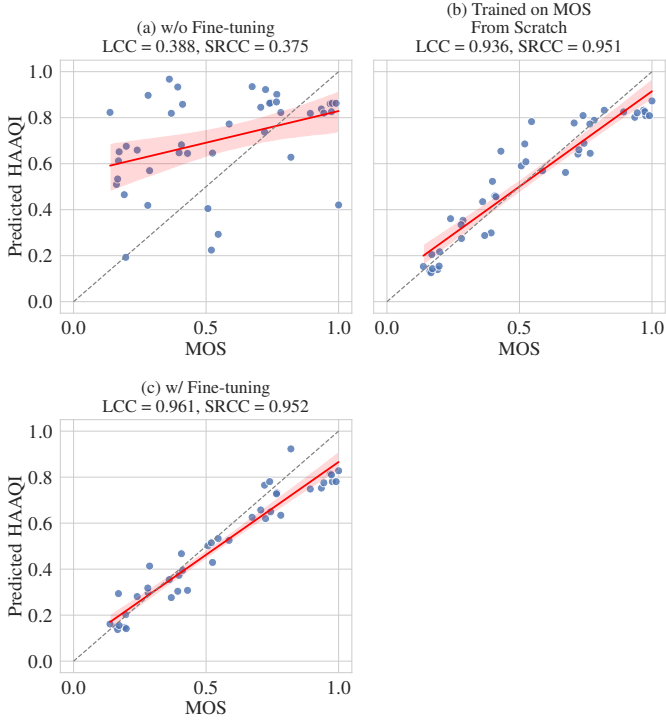}}
    \caption{Scatter plots of Mean Opinion Score (MOS) and HAAQI-Net's predicted HAAQI scores for different conditions. The dashed diagonal line represents the optimal prediction, while the red line represents the regression line with 95\% confidence interval for the model predictions.}
  \label{fig:performance_on_MOS}
\end{figure}

Predicting human subjective scores, such as the MOS, presents a critical challenge in audio quality assessment.
HAAQI-Net, originally designed to predict objective music quality metrics for hearing aid users, offers a promising foundation for adapting to the prediction of human subjective scores. To verify this, we conducted additional experiments. Since there is currently no dataset that includes listening scores from hearing aid users, the experiments were conducted using ODAQ: Open Dataset of Audio Quality \cite{Torcoli2023OdaqOD}. ODAQ is a publicly available dataset containing $240$ audio samples and corresponding quality scores obtained from normal-hearing listeners. Hence, our experiments were conducted using a normal-hearing audiogram to reflect the dataset’s design for normal-hearing listeners.  This dataset provides a reliable baseline for assessing the model's ability to predict subjective human perception of audio quality. The results are presented in Fig. \ref{fig:performance_on_MOS}, which contains scatter plots of the MOS and HAAQI-Net's predicted HAAQI scores under different conditions.

In the first experiment, we evaluated the performance of HAAQI-Net by directly predicting MOS without any fine-tuning. From Fig. \ref{fig:performance_on_MOS}(a), the results were suboptimal. Since HAAQI-Net was initially trained on objective speech quality metrics, it was not fully optimized to capture the nuanced patterns of human subjective judgment, leading to less accurate MOS predictions in this initial setup.

In the second experiment, we trained HAAQI-Net from scratch, using MOS as the target variable. From Fig. \ref{fig:performance_on_MOS}(b), this approach resulted in a significant improvement over the first experiment. By training the model directly on MOS data, HAAQI-Net was able to learn the specific features and patterns that correlate with human subjective perception, leading to better performance. This highlights the importance of aligning the model’s objective function with the target task—in this case, MOS prediction—to capture the nuances of human perception more effectively.

Finally, in the third experiment, we fine-tuned a pre-trained version of HAAQI-Net on MOS data. This fine-tuning approach produced the best results, surpassing the model trained from scratch, as shown in Fig. \ref{fig:performance_on_MOS}(c). Fine-tuning allows the model to retain useful features learned from its original training while adapting them to the specific task of MOS prediction. By leveraging the knowledge gained during pre-training, the model is able to generalize better and refine its predictions, demonstrating the advantages of transfer learning for this task.

To assess the statistical significance of the improvements, we conducted a t-test to compare the performance of the fine-tuned model with that of the model trained from scratch. The resulting p-value of less than $0.05$ indicates that the improvement achieved through fine-tuning is statistically significant. These findings confirm that fine-tuning HAAQI-Net for MOS prediction provides a clear advantage, enhancing the model’s ability to predict subjective scores with greater accuracy.

\subsubsection{Impact of Sound Pressure Level (SPL) Adjustments on HAAQI-Net}

SPL variations can significantly impact speech intelligibility and audio quality assessment. To evaluate the influence of SPL adjustments, we conducted additional experiments to analyze the performance of the proposed algorithm, HAAQI-Net, and its baseline under varying SPL conditions.

In these experiments, we randomly selected $100$ samples from the HAAQI-Net test set and adjusted their SPL to various levels to assess the model's robustness and accuracy under different SPL conditions. In the absence of calibration data for each audio sample, we assumed the reference signal had an SPL of $65$ dB with an RMS of $1.0$, consistent with the methodology employed in the HAAQI paper \cite{kates2015hearing}. This assumption provided a consistent baseline for all experiments. The relative SPL ($SPL_{\rm relative}$) was calculated using the following equation:

\begin{align}
     SPL_{\rm relative} = SPL_{\rm ref} + 20 \log_{10} \left( \frac{RMS_{\rm signal}}{RMS_{ \rm ref}} \right)
\end{align}

where $SPL_{\rm ref} = 65$ dB and $RMS_{\rm ref} = 1.0$. Adjustments  to the SPL were performed by determining the difference ($\Delta SPL$) between the target SPL ($SPL_{\rm target}$) and the current SPL ($SPL_{\rm current}$):

\begin{align}
     \Delta SPL = SPL_{\rm \rm target} - SPL_{\rm current}
\end{align}

This difference was used to compute the gain factor applied to the signal:
\begin{align}
     {\rm Gain Factor} = 10^{\frac{\Delta SPL}{20}}
\end{align}
Although the true reference SPL may vary across audio samples, the assumption of a fixed 65 dB SPL provided a consistent foundation for evaluating SPL adjustments.

The effect of SPL variations on the performance of HAAQI-Net and its baseline was analyzed under three specific conditions:

\begin{itemize}
    \item All SPL Adjustments Combined: Encompasses the full range of SPL variations, including increases and decreases relative to the 65 dB SPL reference.
    \item Lower SPL Adjustments: Examines SPL levels below the reference (35 dB, 45 dB, and 55 dB).
    \item Higher SPL Adjustments: Focuses on SPL levels above the reference (75 dB, 85 dB, and 95 dB).
\end{itemize}

\begin{figure}[t]
    \centering
    \resizebox{\linewidth}{!}{\input{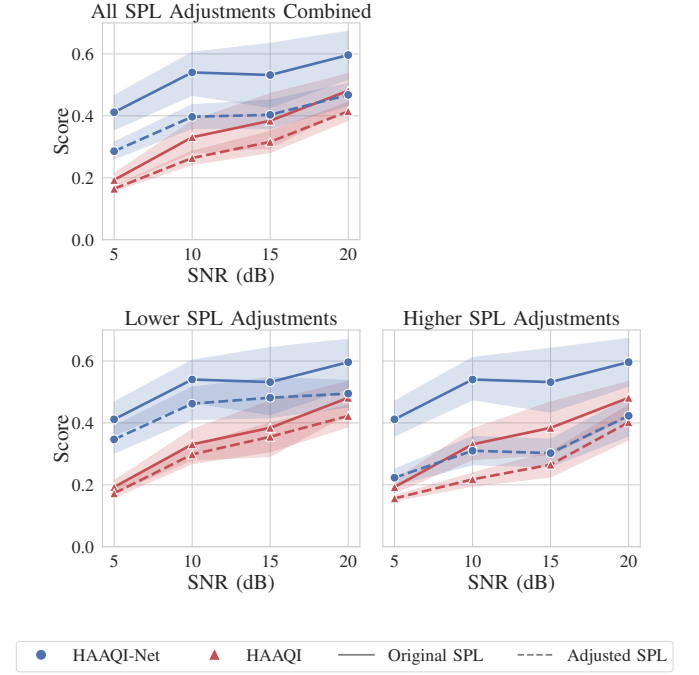}}
    \caption{The performance comparison of HAAQI-Net (blue) and HAAQI (red) under different Signal-to-Noise Ratio (SNR) conditions and Sound Pressure Level (SPL) adjustments.}
    \label{fig:SPL_adjustment_test}
\end{figure}

The corresponding results are visualized in Fig. \ref{fig:SPL_adjustment_test}, where the performance of HAAQI-Net and its baseline is compared across these scenarios. The plots underscore the influence of SPL variations on the model's predictions.

The experimental results demonstrate that HAAQI-Net achieves peak performance when the sound pressure level (SPL) corresponds to the reference value of 65 dB SPL. Deviations from this reference affect the evaluation scores; reducing the SPL below 65 dB SPL leads to a moderate decline in performance, particularly in low signal-to-noise ratio (SNR) conditions, where diminished signal energy increases the challenge of accurate audio quality assessment. In contrast, increasing the SPL above 65 dB SPL results in a more pronounced performance degradation, suggesting that higher SPLs introduce greater distortion or masking effects compared to reductions below the reference level.

These findings align with the human auditory system's behavior, where intelligibility improves with SPL up to an optimal point before declining due to distortion or masking effects. The consistency in trends suggests that the 65 dB SPL assumption is a reasonable anchor for relative SPL measurements and adjustments.

While real-world applications might involve variable SPL conditions, our experiments highlight that HAAQI-Net can maintain reliable performance across a range of SPL levels when adjusted relative to the assumed reference. 

\subsubsection{Comparison With Other Audio Quality Assessment Methods}

This section compares the performance of HAAQI-Net with other audio quality assessment models, including the non-intrusive methods MOSA-Net\cite{zezario2022deep} and HASA-Net\cite{chiang2021hasa}, as well as the intrusive metric PEAQ\cite{thiede2000peaq}. The objective was to evaluate how well HAAQI-Net performs under challenging conditions compared to other methods. The evaluation also assessed its generalizability to normal-hearing listeners, as not all comparison models are tailored for hearing-impaired users.

The experiments were conducted using an unseen music dataset, MUSDB, corrupted with previously unseen noise to test the robustness of the models. To provide a fair basis for comparison, all metrics were normalized to the same scale as HAAQI-Net, with the x-axis representing HAAQI-based anchor points. The evaluation was performed with normal-hearing listeners to eliminate any bias related to hearing aid-specific design and to focus on general performance.

\begin{figure}[t]
    \centering
    \resizebox{\linewidth}{!}{\input{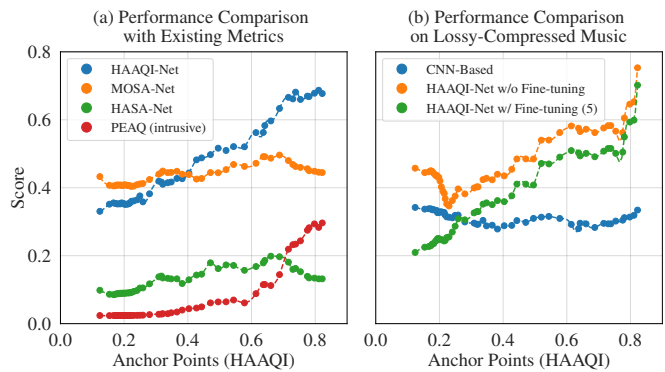}}
    \caption{(a) Comparison of average predicted scores for different methods, including HAAQI-Net, MOSA-Net, HASA-Net, and the intrusive PEAQ metric, across anchor points derived from ground-truth HAAQI scores. (b) Comparison of average predicted scores for different methods, including CNN-Based, HAAQI-Net w/o Fine-tuning, and HAAQI-Net w/ Fine-tuning with 5 samples, across anchor points derived from ground-truth HAAQI scores. Anchor points were calculated as quantiles of the true HAAQI distribution, and model predictions were averaged within a ±0.05 tolerance around each anchor point.}
    \label{fig:compare_with_other_methods}
\end{figure}

The results are presented in Fig. \ref{fig:compare_with_other_methods}(a), which plots the average predicted scores of each model against the anchor points, derived by partitioning the ground-truth HAAQI scores into quantiles. The results indicate that HAAQI-Net performs well across all anchor points. It shows steady improvements as anchor points increase, often outperforming HASA-Net and matching or exceeding MOSA-Net, particularly at higher anchor points. HASA-Net's performance was the lowest overall, while MOSA-Net performed competitively in the mid-range but started to decline at higher anchor points. The intrusive PEAQ metric scored the lowest among all methods, highlighting its limitations in unseen noisy conditions. These findings suggest that while HAAQI-Net was designed with hearing aid users in mind, it remains effective in broader contexts for normal-hearing listeners.

We further evaluate HAAQI-Net's performance on a specific application: assessing the quality of lossy-compressed music recordings\cite{cnn_for_lossy}. This experiment also includes a comparison with the CNN-based method proposed in \cite{cnn_for_lossy}. The experiment focuses on the task of assessing audio quality in lossy-compressed music, a domain where effective non-intrusive solutions are essential.

The results are presented in Fig. \ref{fig:compare_with_other_methods}(b), which plots the average predicted scores of each model against the anchor points, derived by partitioning the ground-truth HAAQI scores into quantiles. In the initial setup, HAAQI-Net was applied directly to the lossy-compressed music data without any fine-tuning. The results showed that HAAQI-Net outperformed the CNN-based approach, particularly in the mid-to-high score range. However, it faced challenges in accurately predicting the lower score region, indicating limitations in generalization for such cases. To address this, we fine-tuned HAAQI-Net using a very small dataset comprising only five samples, demonstrating a few-shot learning approach.

Fine-tuning significantly improved the performance of HAAQI-Net in the lower score range, as observed in the plot. This improvement highlights the adaptability of HAAQI-Net to new tasks with minimal additional data. This few-shot learning capability is a key advantage, as it allows HAAQI-Net to be effectively tailored to specific applications with minimal effort, broadening its usability across diverse audio quality assessment tasks.

\subsubsection{Efficiency Evaluation} 
\begin{figure}[t]
    \centering
    \includegraphics[width=\linewidth]{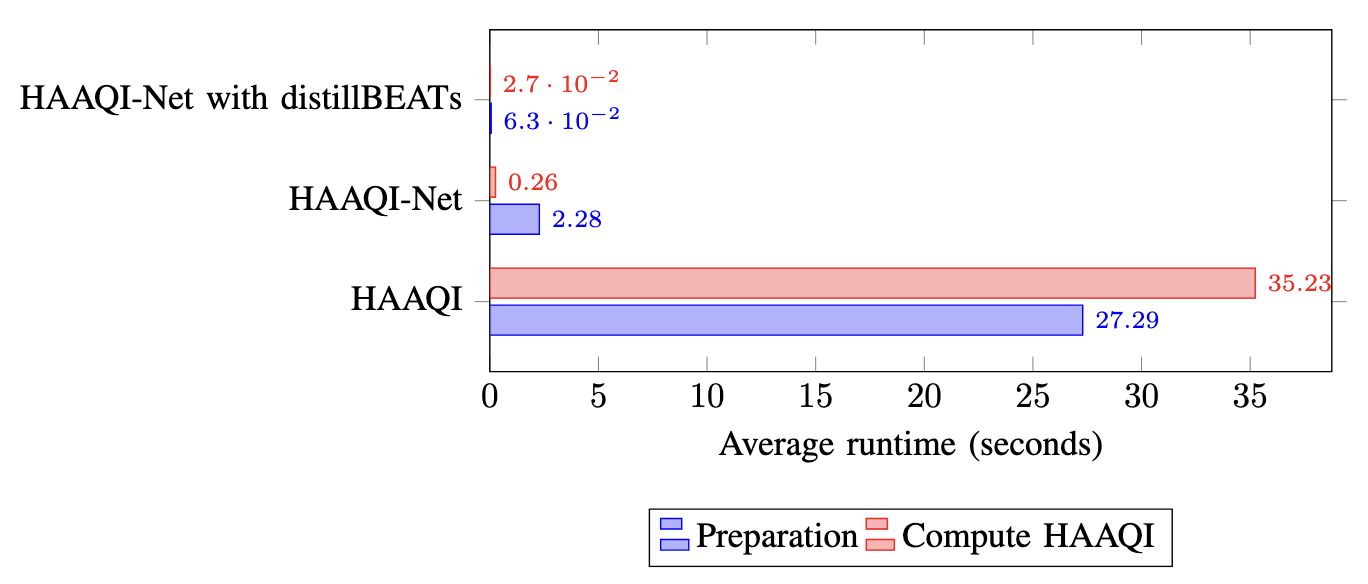}
\caption{Comparison of the runtimes of HAAQI, HAAQI-Net, and HAAQI-Net with distillBEATs\&ws.}
  \label{fig:runtime_comparison}
\end{figure}

Fig. \ref{fig:runtime_comparison} provides a comprehensive comparison of the runtimes of different stages of HAAQI, HAAQI-Net, and HAAQI-Net with distillBEATs. 

For the conventional HAAQI method, the pre-processing stage involves extensive computation, resulting in an average runtime of $27.29$ seconds per audio. This time-consuming process mainly includes feature extraction and signal processing tasks necessary to derive relevant audio features. In comparison, HAAQI-Net significantly reduces pre-processing time, with an average feature extraction time of only $2.28$ seconds per audio. This substantial improvement can be attributed to HAAQI-Net's efficient neural network architecture, which uses BEATs for feature extraction, effectively reducing computational overhead.
In addition, HAAQI-Net achieves remarkable speed enhancements in HAAQI score prediction, requiring only $0.26$ seconds on average per audio, compared to $35.23$ seconds required by the conventional HAAQI method. This efficiency results from HAAQI-Net's ability to efficiently process extracted features and use optimized neural network layers and computational strategies to produce accurate predictions. Overall, HAAQI-Net exhibits outstanding efficiency, taking only $2.54$ seconds per audio to complete the entire prediction process, while the conventional HAAQI method requires $62.52$ seconds. This remarkable performance highlights the effectiveness of deep learning techniques in accelerating music quality prediction tasks while maintaining high accuracy.


Furthermore, additional efficiency improvements can be achieved by integrating distillBEATs into HAAQI-Net. The runtime for pre-processing is further reduced to just $0.063$ seconds per audio, while the calculation of HAAQI scores takes an astonishing average of $0.027$ seconds. This significant improvement is attributed to the distillation process, which enhances model efficiency by transferring knowledge from a larger pre-trained model (BEATs) to a smaller target model (distillBEATs). As a result, HAAQI-Net with distillBEATs achieves unparalleled speed and scalability, making it well-suited for real-time music audio quality assessment applications where rapid processing is crucial.

These thorough evaluation results highlight the transformative impact of HAAQI-Net in revolutionizing music quality prediction, offering unprecedented efficiency and performance compared to traditional methods.



\section{Conclusions}
The development and evaluation of HAAQI-Net mark significant progress in music audio quality assessment for hearing aid users. Through systematic experimentation, our research demonstrates HAAQI-Net's effectiveness, versatility, and efficiency across various scenarios and datasets. Our investigation into different input features reveals that pre-trained SSL models like WavLM and BEATs, particularly when paired with adapters, excel in capturing acoustic characteristics and semantic nuances, ensuring superior performance in music quality prediction.

Evaluations of HAAQI-Net's generalization ability on seen and unseen datasets highlight its consistent performance, despite challenges such as overfitting with certain feature representations. The BEATs model consistently outperforms other models, effectively capturing and encoding general aspects of music quality. Additionally, HAAQI-Net's adaptability across varied hearing-loss patterns and processing conditions underscores the importance of specialized feature representations in accurate music quality prediction, proving its effectiveness and reliability in different scenarios.

Experiments adapting HAAQI-Net to predict subjective scores, such as MOS, highlight the model’s flexibility and potential to bridge the gap between objective metrics and subjective human perception. Fine-tuning on MOS data achieved significant improvements, demonstrating the advantages of transfer learning for predicting subjective scores with greater accuracy. Similarly, tests under varying SPL conditions reveal HAAQI-Net's robustness, with optimal performance at 65 dB SPL, aligning with human auditory system behavior. While deviations affect accuracy, the model maintains reliable predictions across a range of SPL levels, showcasing its practical utility in real-world scenarios.

Moreover, HAAQI-Net achieves substantial efficiency improvements, significantly reducing processing time compared to traditional methods while maintaining high accuracy. The integration of distillation strategies further enhances this efficiency, positioning HAAQI-Net with distillBEATs as a promising solution for real-time music audio quality assessment. In summary, HAAQI-Net advances the assessment of music quality for hearing aid users by capturing musical nuances, generalizing across diverse scenarios, and delivering efficient performance, thereby enhancing the listening experience for individuals with hearing impairments. Future research will focus on optimizing HAAQI-Net's performance and exploring its application in real-world settings.

\bibliographystyle{IEEEtran}
\bibliography{IEEEabrv, IEEEexample}

\newpage

\vfill

\end{document}